# Boosting bottom-up milling of glass with GHz laser bursts

Miglė Mackevičiūtė, Juozas Dudutis, Aleksandras Kondratas, Paulius Gečys

Center for Physical Sciences and Technology (FTMC), Savanorių Ave. 231, LT-02300 Vilnius, Lithuania

Corresponding author: migle.mackeviciute@ftmc.lt

**Abstract**

Burst-mode laser processing has gained increasing attention in material machining due to its potential to enhance processing efficiency and quality. However, most studies have focused on conventional top-down machining strategies, while investigations of bottom-up milling (BUM) remain limited. In this work, we optimize processing parameters for high-efficiency glass milling using a bottom-up approach with GHz bursts. The results show that the highest BUM efficiencies are achieved when employing high pulse counts (≥ 20). To the best of our knowledge, we report the highest glass milling efficiency, reaching more than 250 μm$^3$/μJ for both soda-lime glass and fused silica. Furthermore, increasing the burst repetition rate while maintaining a parameter set associated with high milling efficiency enables a material removal rate larger than 8 mm$^3$/s.

## 1. Introduction

The use of high-frequency laser bursts, typically MHz or GHz, has proven to be beneficial in many areas of glass processing, including laser-assisted cutting using non-diffractive beams [1] or laser filamentation [2], chemical etching [3], high-aspect-ratio top-down percussion drilling [4,5], bottom-up cutting [6], and micro-welding [7]. Material excitation by high-frequency bursts advances from the accumulation of heat and defects, created transient material states, and reduced nonlinear detrimental effects, leading to improved energy deposition [1]. Laser bursts also enable unique processing regimes, such as polarization-controlled directional micro-cracks facilitating glass scribing [8,9].

The majority of laser-based glass milling studies are conducted using a top-down milling (TDM) approach, with the beam focused to the top surface of the sample and successive layers of material removed from the exposed surface. The method is simple and versatile, as it can be used for a wide range of transparent or opaque materials. However, the material removal efficiency is limited due to the laser pulse interaction with the plasma and the removed particles [10,11]. The absorption of subsequent pulses by the plasma leads to its expansion toward the laser source, reducing the amount of laser energy reaching the sample [12]. In addition, TDM results in tapered walls due to reduced effective fluence on the sidewall surfaces, arising from enlarged beam spot area and increased Fresnel reflections [13,14]. Furthermore, beam refraction at the V-shaped walls can result in parallel, periodic band-like defects along the bottom surface [15] and spicular volumetric damage near the ablated trenches [16].

Transparent materials with low reflectance and high top-surface quality can be machined by focusing the beam to the bottom surface, with material removal progressing upward through the sample. In bottom-up milling (BUM), beam interaction with the sidewalls is mitigated, enabling vertical, slanted, or horizontal wall profiles [17,18]. BUM is also more energy-efficient than TDM for several reasons. First, the bottom surface exhibits a lower damage threshold than the top surface due to constructive interference of transmitted and reflected light at the glass-air interface [19]. Second, beam interaction with the ablation products is mitigated as gravity and shockwaves facilitate their removal [16]. Third, plasma-heated surrounding material increases beam absorption at the bottom surface [20]. In addition, partially confined superheated material is formed in the bulk of the material as the absorption front shifts toward the laser source [21].

The material removal rate and the quality of the processed surface depend heavily on the laser pulse duration [18,22]. High-quality material processing is typically achieved with femtosecond pulses, whose durations are shorter than picosecond-scale energy relaxation times, thereby confining energy deposition and enabling ablation with a minimal heat-affected zone [23,24]. However, in BUM, beam propagation through the material at high peak intensities can induce nonlinear effects that, under low numerical aperture (NA) focusing, can lead to filamentation with intensity clamping below the threshold required for material modification [25,26]. Self-focusing and intensity clamping effects can be mitigated by using high NA focusing [27]. Increasing numerical aperture improves laser machining precision by reducing the focal spot size and depth of focus, but it also reduces usable scan field size, thereby limiting galvanometer system flexibility.

High-throughput BUM of glass can be achieved by using nanosecond pulses, which induce strong thermal stresses that promote material cracking [18]. Since cracking requires less energy than ablation [20], significantly higher processing rates are attainable. Moreover, longer pulse durations reduce peak intensity and mitigate nonlinear propagation effects such as self-focusing and intensity clamping, enabling more efficient energy deposition. This allows the use of low-NA focusing, resulting in a larger focal spot, an extended depth of focus, a larger scan field size and speed. Although this approach comes at the expense of spatial precision, it enables high-throughput machining over larger areas. However, longer pulses result in lower peak intensities and, consequently, weaker nonlinear absorption. Thus, shorter pulses tend to enhance nonlinear absorption, whereas longer pulses can improve energy deposition by mitigating self-focusing and the associated intensity clamping. To enhance nonlinear absorption while exploiting the advantages of longer pulses, second-harmonic radiation can be employed [18]. However, this approach introduces trade-offs, including losses from second-harmonic conversion and reliability concerns such as nonlinear crystal damage, both of which can reduce the available laser output power.

Burst-mode irradiation has also been shown to improve material removal efficiency through crack-driven mechanisms [28]. Furthermore, it was demonstrated that the bottom-up cutting of soda-lime glass using GHz bursts of picosecond pulses at 1030 nm resulted in slightly higher cutting efficiency compared with nanosecond pulses at 532 nm [6]. Because the material removal mechanism is strongly influenced by the number of pulses per burst, tuneable burst length provides an additional degree of freedom for controlling the ablation process. Despite recent advances, BUM of glass using GHz bursts remains largely unexplored. In this study, we systematically investigate the laser processing conditions for high-throughput bottom-up milling of soda-lime glass and fused silica using GHz bursts.

## 2. Materials and methods

Properties of the laser setup are listed in Table 1. Experiments were conducted with a laser operating at the central wavelength of 1030 nm. The circularly polarized laser beam was guided in the XY plane by galvanometer scanners and focused by the telecentric f-theta lens [Figure 1(a)]. The vertical (Z) position was adjusted by moving the galvanometric scanners and the focusing lens.

Table 1. Properties of a laser setup.

| **Parameter/ Tool** | **Value/ Name** |
|---|---|
| Laser | Carbide CB3-80W (Light Conversion) |
| Central wavelength ($\lambda$) | 1030 nm |
| Pulse duration ($\tau_p$) | 5 ps |
| Burst repetition rate ($f_b$) | 100, 200, 400 kHz |
| Intra-burst repetition rate ($f_i$) | 2.2 GHz (465 ps) |
| Pulses per GHz burst ($N$) | 10; 15; 20; 25; 30 |
| Polarization | Circular |
| Galvanometer scanners | intelliSCAN 14 (Scanlab) |
| Focal length of the telecentric f-theta lens | 100 mm |
| Numerical Aperture (NA) | 0.04 |
| Beam waist diameter (measured by Liu's method at the 1/e$^2$ intensity level [29]) | 18 µm |
| Z-axis positioner | PRO165LM (Aerotech) |

The laser could operate in the GHz burst regime emitting pulses with the temporal separation of 465 ps (2.2 GHz). The temporal characteristics of bursts were measured by the detector D-15ir (Picometrix) and oscilloscope DPO72004C (Tektronix). It should be noted that the employed number of pulses per burst up to 30 exceeded the standard limit of the laser (10), and some additional pulses (pre-burst) appeared at large burst counts (≥ 20), which were otherwise absent in the standard regime. The oscillograms of the GHz burst regimes are shown in Figure S1 in the supplementary material. For the sake of clarity, the burst regime will be denoted only by the nominal number of pulses in a burst.

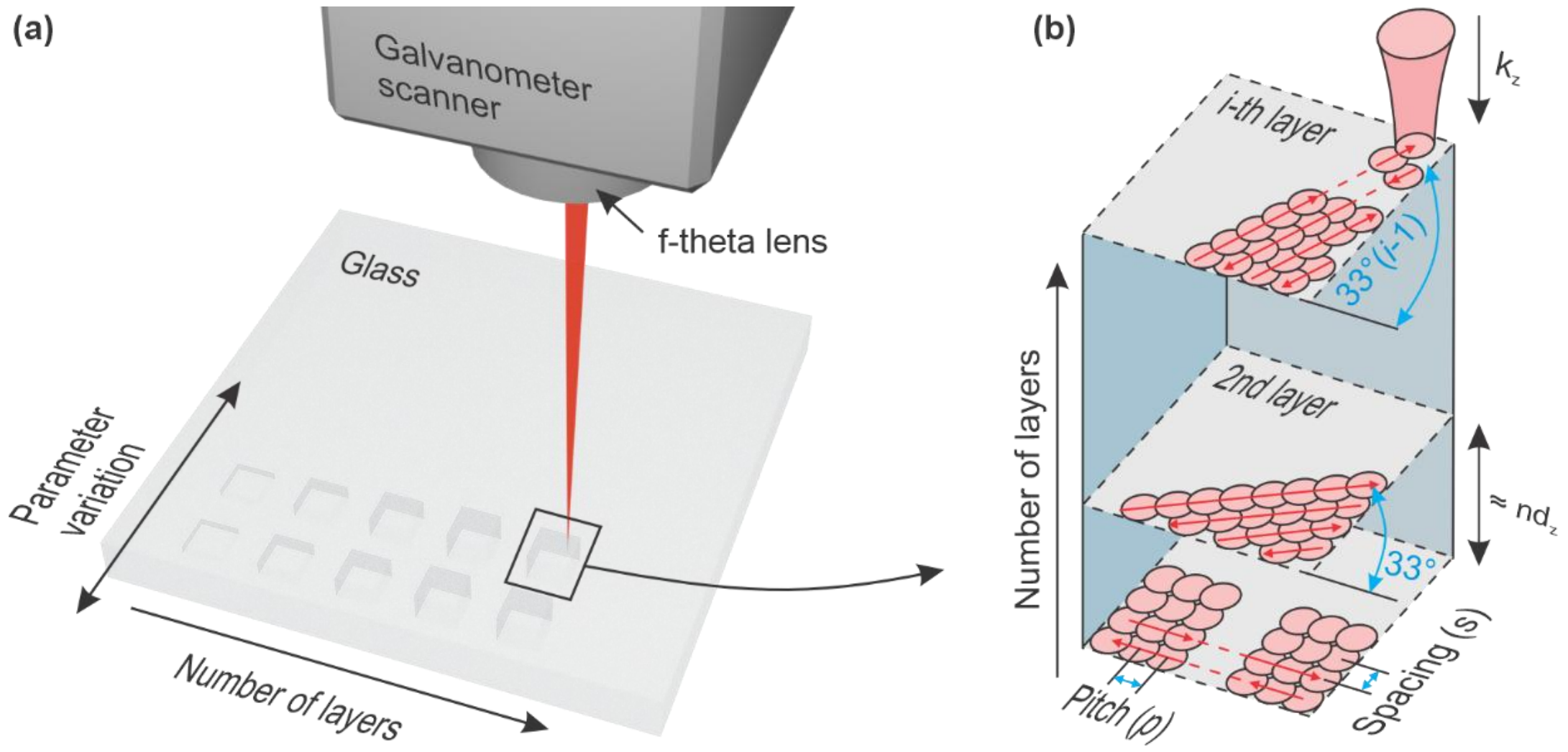

Figure 1. Experimental setup. **(a)** Milling of cavities with an increasing number of layers. **(b)** Laser beam scanning algorithm in the transverse plane.

Unless otherwise specified, the applied burst repetition rate ($f_b$) was 100 kHz, which corresponds to the time interval of 10 µs between bursts. To reduce nonlinear propagation effects, such as light filamentation, which are significant at weak focusing and disrupt energy delivery in the volume of material, the laser pulse duration was increased to 5 ps, gaining a negative chirp.

The role of burst energy and number of pulses per GHz burst was investigated. The experiment was conducted by milling 3 × 3 $mm^2$ squares in two materials, soda-lime glass (SLG) and fused silica (FS), with thicknesses of 4.8 mm and 6.3 mm, respectively. Cavities were milled using a different number of layers applying the bi-directional scanning technique, which was rotated by 33° after each layer to obtain a more homogeneous overlap [Figure 1(b)]. The distance between subsequent shots is denoted as pitch ($p$), and the distance between the subsequent lines (hatch) is denoted as spacing ($s$). The number of layers was varied from 1 to 30, and the increment between successive cavities was varied from 1 to 5. Initially, the geometric focus position was set below the rear surface to avoid the instantaneous failure of the glass due to induced intra-volume modifications and subsequent laser-assisted growth of cracks. The focal position was gradually moved upwards between successive layers to maintain quasi-constant fluence in the processing area. Following the small-angle approximation, the focal shift in glass is $\approx nd_z$, where $n$ is the material refractive index, and $d_z$ is the shift of the beam waist position in air.

The depth and surface roughness of milled cavities were measured with the optical profilometer S Neox (Sensofar). Surface roughness Sa of the bottom of milled cavities was measured over a 2 × 2 $mm^2$ area using a 20× objective (NA = 0.45). The bottom surface chippings were evaluated by measuring the maximum chip length of the milled cavities using an Eclipse LV100NDA (Nikon) optical microscope with a 5× objective (NA = 0.15). For 3–4 cavities, the roughness and length of the longest chippings were averaged separately and reported as (mean ± SD), with the standard deviation (SD) calculated from the measured data.

JSM-IT510 (JEOL) scanning electron microscope (SEM) was used to investigate the surfaces of the machined samples. Samples were additionally coated with a 25 nm-thick layer of silver using a sputter coater Q150T ES (Quorum Technologies) to enhance the SEM imaging.

## 3. Results and discussion

*Milling evolution*

Before milling each cavity, the laser beam was focused below the rear surface of a sample. In BUM, the first-layer scan is performed on an initially undamaged surface, whereas subsequent scans interact with the modified, roughened surface generated by the preceding laser passes. Roughened surface has defects, which change absorption bands [30], cause light intensification due to (i) local enhancements of the electric field [31] or (ii) interference between incident and reflected light [32,33], reduce material strength [34], and thereby facilitate further processing. Consequently, the optimal processing parameters differ between the initial and subsequent laser scans. For high-bandgap, low-defect materials, additional surface roughening might be required to achieve optimal shot-to-shot spatial overlap. Once the processing area is uniformly modified and material removal becomes stable, the milled depth increases linearly with the number of processed layers [Figure 2 (a)].

Initial surface roughening was necessary to achieve fast milling of fused silica. In this step, 6 layers were scanned using the single-pulse regime with the pulse energy of 50 μJ, $\tau_p$ = 5 ps, $p$ = 3 μm, $s$ = 18 μm, and $d_z$ = 20 μm. The initially milled cavities had an average depth of 50 μm, a surface roughness of Sa = (2.6 ± 0.2) μm, and surface chippings of (15 ± 4) μm. In contrast, soda-lime glass was processed without pre-roughening, owing to its lower bandgap, which facilitates free-electron plasma generation in the pristine material [35,36]. For high-throughput milling, the same pitch and spacing were used for both fused silica and soda-lime glass ($p$ = 12 μm, $s$ = 18 μm).

The shift in the beam focus position between layers played a strong role in the further evolution of milled cavities. If it was too low, the material removal rate was limited due to increasing beam waist diameter and decreasing fluence with increasing cavity depth [Figure 2 (b)]. Notably, such self-terminated mechanism can be used to control the thickness of the removed material per layer ($h \approx nd_z$) and increase the processing resolution along the vertical axis. When $nd_z$ was larger than the removed depth per layer, the high-intensity beam waist started to penetrate the material at some point, which led to (i) the uncontrollable in-volume crack growth and defects or (ii) the beam waist position moving away from the processing zone, resulting in the reduced fluence and complete termination of milling. Severe cracking of glass was observed at high burst energies, while the second scenario most likely occurred at low burst energies. Additionally, the depth began to deviate from a linear dependence at some point. Thus, the optimization aimed to find the optimal shift in waist position to avoid generation of defects while maintaining high throughput.

The slope of the linear fit of the milled cavity depth versus the number of layers represents the depth removed per layer ($h$) and was used to calculate the material removal rate and milling efficiency. The standard error of the fitted slope was used to calculate the corresponding uncertainties of the material removal rate and milling efficiency. These uncertainties are reported as error bars in the following text and figures.

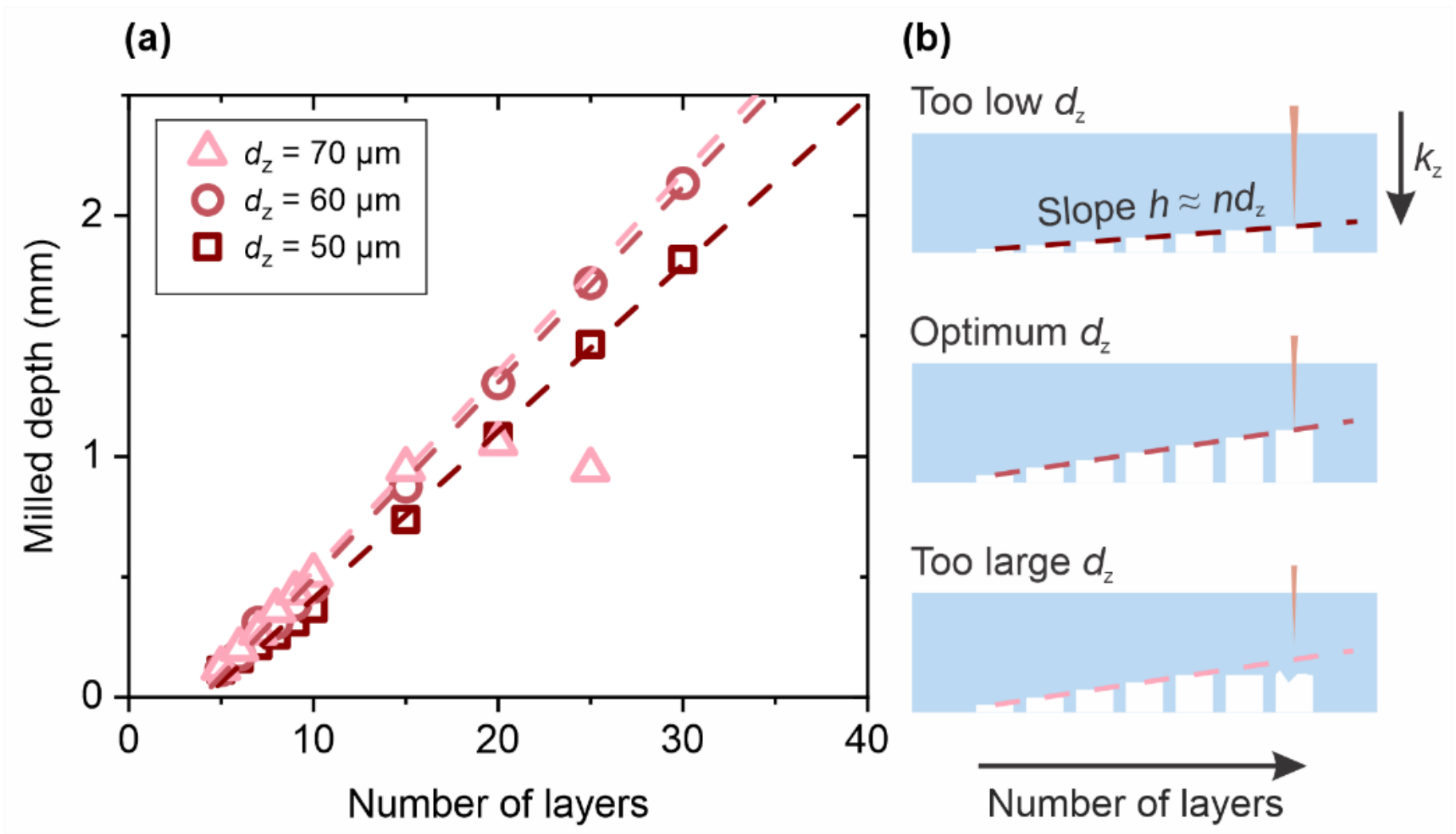

Figure 2. Evolution of cavities. **(a)** Milled depth in soda-lime glass versus number of layers for different mechanical focal shift using burst energy of 78 µJ and 25 pulses per burst. Dashed lines show linear fits. **(b)** Schematic illustration of changes in cross-section of milled cavities at different focal shifts.

*Milling throughput*

Milling throughput was evaluated for different number of pulses per GHz burst (10–30) and burst energies (≈ 35–150 µJ). The applied burst repetition rate was 100 kHz. Figure 3(a, b) shows the milled depth per layer ($h$). The relative errors reached 8% and 19% for soda-lime glass and fused silica, respectively, as indicated by the standard errors of the fitted slopes. For some burst counts, after a swift increase with increasing burst energy, the depth per layer saturated or even slightly decreased at the maximum energy. The largest values were achieved at the burst count of 20 and were over 90 µm and 110 µm for fused silica and soda-lime glass, respectively. It corresponded to the theoretical material removal rate (MRR) of around 2 $mm^3$/s or larger, estimated according to:

$$\mathrm{MRR} = pshf_b, \quad (1)$$

where $p$ is pitch, $s$ – spacing, $h$ – removed depth per layer, and $f_b$ – burst repetition rate. It does not consider the time wasted on jumps and scanning delays; thus, the actual processing rate is always lower.

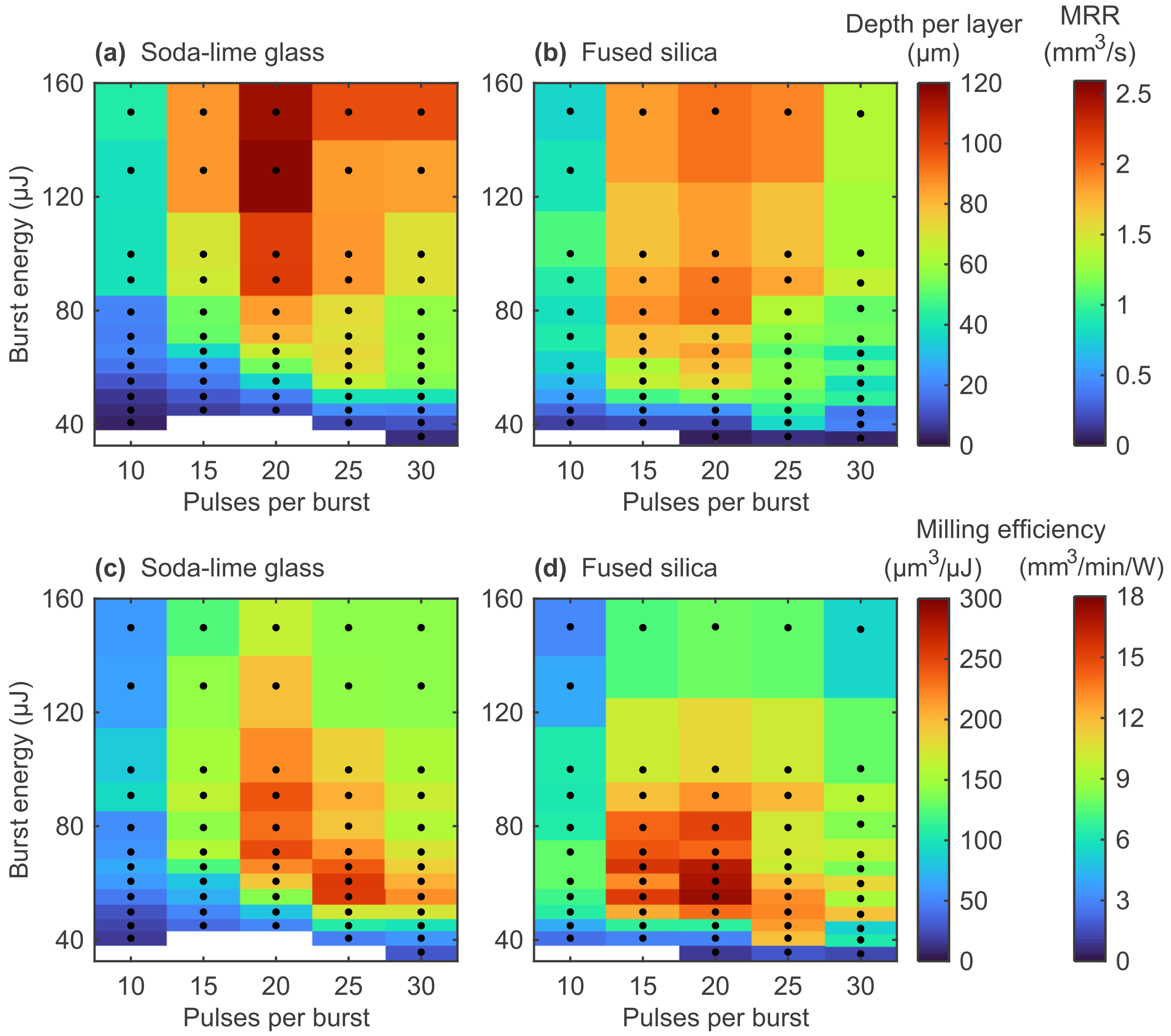

Figure 3. Milling throughput. **(a, b)** Milled depth per layer and material removal rate and **(c, d)** milling efficiency of **(a, c)** soda-lime glass and **(b, d)** fused silica versus burst parameters. Solid black dots show experimental data points. Nearest neighbour interpolation was used to generate 2D maps.

Figure 3(c, d) depicts milling efficiency, which was estimated as:

$$\eta = psh/E_b\,, \tag{2}$$

where $E_b$ is the net energy of the burst, including the pre-burst pulse, if present. The largest milling efficiency of soda-lime glass was (256 ± 2) μm$^3$/μJ at 25 pulses per burst. In general, efficiency over 200 μm$^3$/μJ could be obtained at 20–30 pulses per burst with the material removal rate of 1.2–2.2 mm$^3$/s.

The milling efficiency of fused silica was slightly larger, with the maximum value of (288 ± 2) μm$^3$/μJ achieved at 20 pulses per burst. The efficiency larger than 200 μm$^3$/μJ could be obtained at 15–25 pulses per burst with the material removal rate of 1–2 mm$^3$/s.

*Milling quality*

The milled surface area roughness Sa and the maximum chipping length dependence on burst energy and pulses per burst are shown in Figure 4(a, b) and Figure 4(c, d), respectively. For this, 3–4 cavities, corresponding to the stable (linear) milling regime, were evaluated. Other processing parameters are the same as of regimes shown in Figure 3.

The average roughness Sa was in the range from 3 µm to 10 µm with the relative standard deviation of up to 25%. Under the most efficient milling conditions, Sa was (5.8 ± 0.2) µm and (7.5 ± 0.4) µm for soda-lime glass and fused silica, respectively. Higher-quality cavities with a roughness of ≈ 5 µm could be milled at a lower number of pulses per burst, resulting in a moderate milling efficiency between 100–200 µm$^3$/µJ.

In most cases, the maximum chipping length was larger for fused silica and ranged from 43 to 170 µm, while for soda-lime glass it ranged from 25 to 95 µm. However, under the most efficient milling conditions, the maximum chipping lengths were comparable, measuring (56 ± 9) µm for soda-lime glass and (54 ± 6) µm for fused silica.

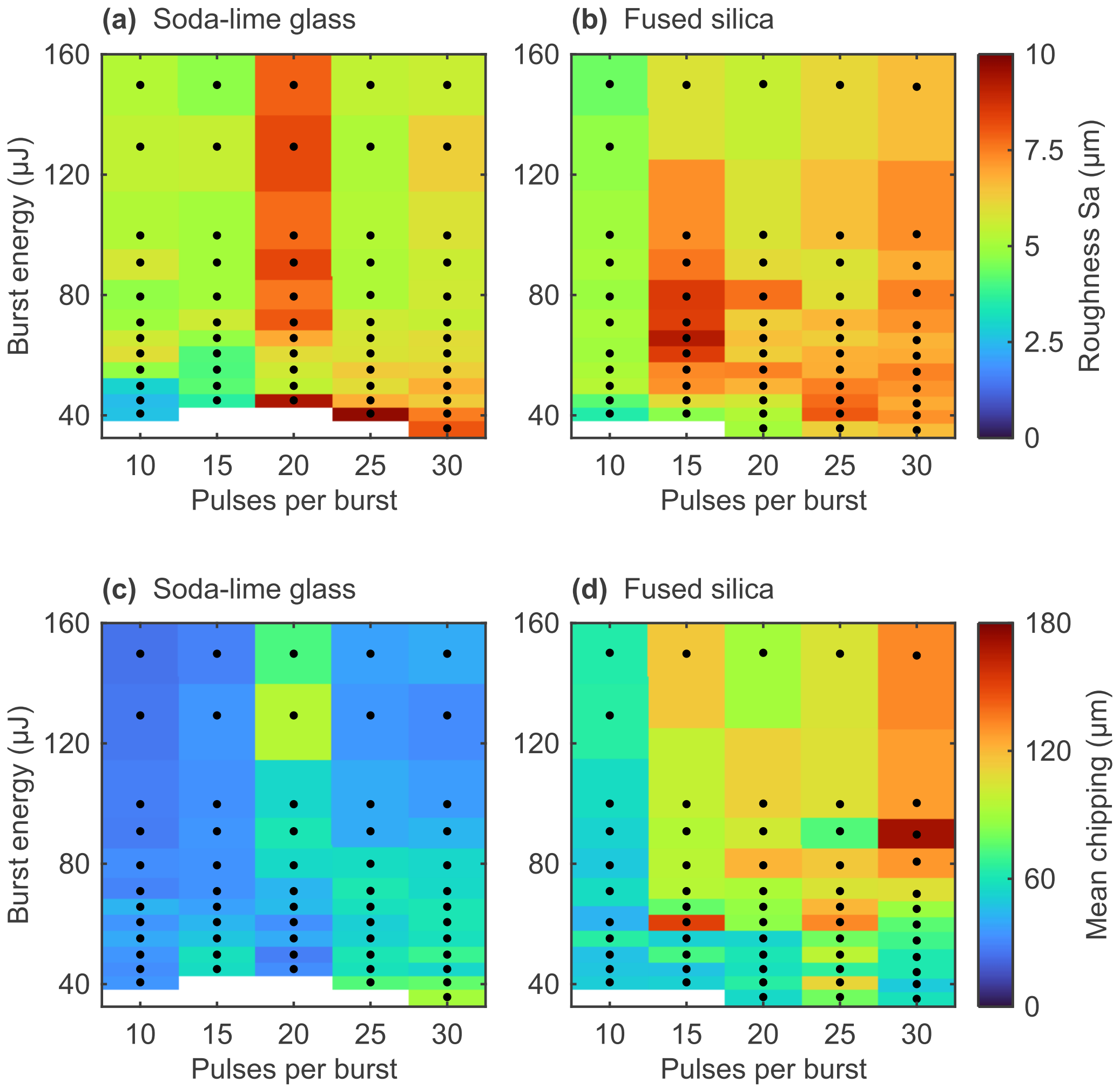

Figure 4. Milling quality using GHz bursts. **(a, b)** Roughness Sa of the milled bottom surface and **(c, d)** maximum surface chippings of **(a, c)** soda-lime glass and **(b, d)** fused silica versus burst parameters. Solid black dots show experimental data points. Nearest neighbour interpolation was used to create 2D maps.

The parameter maps in Figure 3 and Figure 4 suggest that milling efficiency and defect formation are related. The variation in surface quality parameters with milling efficiency is illustrated in Supplementary Figure S2. A clear trend emerges, with higher milling efficiency generally associated with increased surface roughness, indicating a trade-off between efficient material removal and surface quality. However, the highest milling efficiency does not necessarily correspond to the most pronounced damage, suggesting that appropriate combinations of burst energy and pulse number can maintain high efficiency while mitigating defects. This is particularly evident in fused silica, for which the highest milling efficiency is achieved at 20 pulses per burst and burst energies of 50–70 µJ [Figure 3(d)], whereas the highest surface roughness occurs at approximately 15 pulses per burst under comparable energies [Figure 4(b)].

The milled squares produced under the most efficient milling conditions were cut in half to expose their cross-sections for SEM examination. The resulting cut surface was subsequently mechanically polished. The SEM micrographs of the surfaces are shown in Figure 5. The 3D micrographs reveal that the milled squares exhibit vertical sidewalls. The processed surface morphology indicates that material removal was dominated by crack-induced fracture, as demonstrated by cleaved features spanning several tens of micrometres. This material removal mechanism accounts for the high milling efficiency observed.

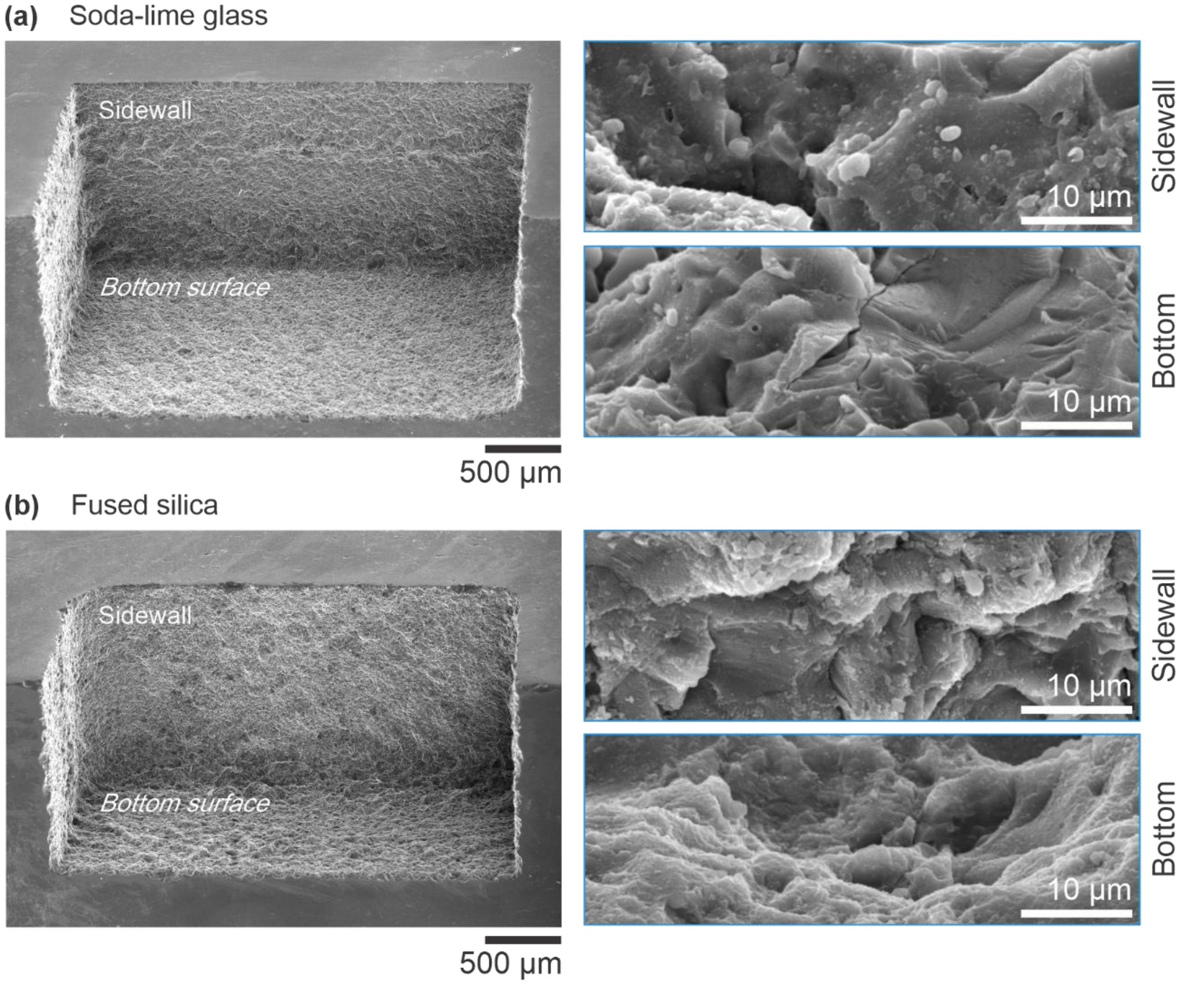

Figure 5. Surfaces milled with GHz bursts. SEM micrographs of **(a)** soda-lime glass and **(b)** fused silica processed using the parameters that resulted in the highest milling efficiency. The samples were tilted at 30°.

In addition, sub-micrometre to micrometre-sized structures are observed on both the sidewalls and bottom surface, with a higher density on the vertical sidewalls. These features generally exhibit relatively smooth, rounded boundaries and appear to be attached to the processed surface. Their morphology and distribution may be indicative of redeposition and/or resolidification of material generated during laser ablation. The bottom surface contained fewer of these features compared to the sidewalls. This might be due to gravity and induced shockwaves facilitating the removal of ablation products from the bottom surface, directing them downwards and laterally toward the surrounding sidewalls. Furthermore, gaseous material, liquid, partially melted particles, and fractured debris may be ejected at different timescales. Time-resolved imaging of the rear-side of fused silica excited by nanosecond pulses indicates that fractured particles are emitted from the material last [21]. Thus, less molten material should remain on the bottom surface.

*Scaling-up the throughput*

Additionally, milling of soda-lime glass was investigated at 100, 200, and 400 kHz burst repetition rates [Figure 6(a)]. The parameter set of 20 pulses per burst and $E_b$ = 71 µJ was selected, yielding a high milling efficiency of (247 ± 5) µm³/µJ at 100 kHz [Figure 3(c)]. The material removal rate scales linearly with burst repetition rate, increasing fourfold from 100 to 400 kHz and reaching (6.1 ± 0.1) mm³/s [Figure 6(a)]. With constant pitch, increasing the burst repetition rate directly increases the scanning speed. Thus, once the optimal burst energy is established, throughput can be increased by raising the repetition rate, up to limits set by laser power or scanner speed.

The milling efficiency and rate dependence on burst energy were investigated for 100 and 400 kHz burst repetition rates [Figure 6(b, c)]. The milling efficiency increased with burst energy up to 60–70 µJ, after which it began to decrease. Up to 60 µJ, both burst repetition rates yielded comparable milling efficiencies. However, above 70 µJ, 100 kHz yielded a higher milling efficiency compared with 400 kHz. Nevertheless, operating at 400 kHz enables a fourfold increase in scanning speed, resulting in a maximum material removal rate of (8.49 ± 0.03) mm$^3$/s at a burst energy of 137 µJ, while maintaining a moderate surface roughness of Sa = (5.6 ± 0.5) µm. Therefore, a fourfold increase in the burst repetition rate allowed for achieving 3.3 times higher milling rate.

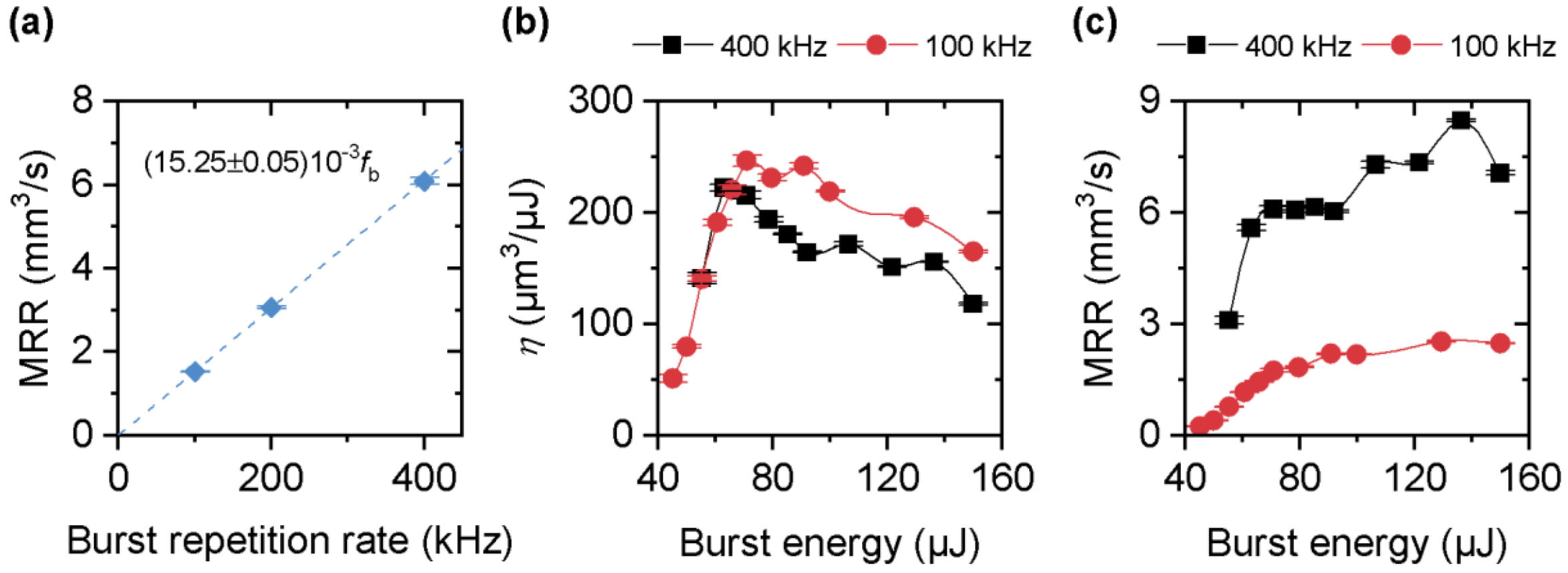

Figure 6. Enhancing throughput of soda-lime glass milling via burst repetition rate. **(a)** Material removal rate dependence on burst repetition rate when using 71 µJ burst energy and 20 pulses per GHz burst. A dashed line depicts a linear fit. **(b)** Milling efficiency and **(c)** material removal rate dependence on burst energy when using 100 and 400 kHz burst repetition rates. Solid lines in **(b)** and **(c)** are for eye guidance. The error bars were calculated from the standard errors of the fitted slopes obtained from the linear fits of milled cavity depth versus the number of processed layers.

*Applications*

To demonstrate the flexibility of the technology, BUM has been applied for the fabrication of complex structure (a 25 × 32 × 3.4 $mm^3$ coat of arms of Lithuania) and small diameter channels [Figure 7]. The complex structure was formed from 6.3 mm-thick fused silica within a total processing time of ~1 h, including a 10 min initiation phase and 49 min of milling. The processing was carried out using 20 pulses per burst, $E_b$ = 150 µJ, and $f_b$ = 100 kHz, corresponding to the highest material removal rate regime at this burst repetition rate [Figure 3(a)]. Under these conditions, a theoretical material removal rate of 1.25 $mm^3$/s was achieved.

The material removal rate was lower than that obtained during process optimization because the focus position shift per layer had to be reduced. For large-area machining, variations in the maximum achievable machining depth may arise from residual optical deviations of the f-theta lens. Furthermore, during machining of larger areas, the longer interval before a given location is reprocessed allows greater heat dissipation, which can potentially result in a lower machined depth per layer. Consequently, the material removal depth at certain locations may be less than $nd_z$, potentially leading to crack formation or premature termination of the milling process. Selecting a smaller $d_z$ mitigates these effects ensuring more uniform material removal across the entire machined area.

The milling of the channels was demonstrated in a 4.8 mm soda-lime glass sample using 20 pulses per burst, $E_b$ = 130 µJ, and $f_b$ = 5 kHz. The burst repetition rate was reduced to lower the scanning speed and minimize thermal accumulation. This resulted in a lower theoretical material removal rate of 0.02 $mm^3$/s. Channels with diameters as small as 200 µm were successfully formed. However, at 200 µm, dust particles were observed to fuse to the channel walls. Clean channels, without visible dust deposits, were obtained for diameters of 400 µm and larger. The chippings were up to 60 µm for both surfaces.

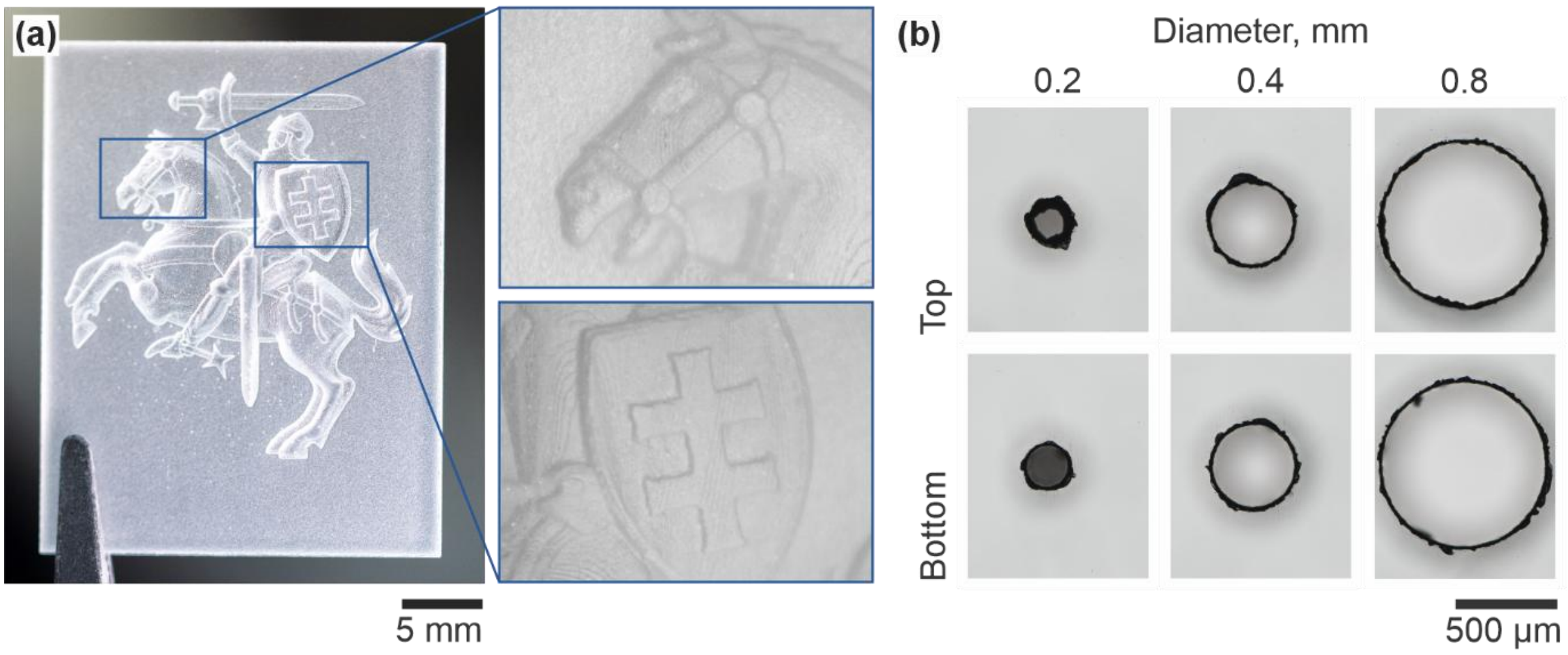

Figure 7. Complex milled structures. **(a)** Coat of arms of Lithuania (Vytis). Photo was taken by Valdas Jarutis. **(b)** Optical microscope pictures of the top and bottom surfaces of the milled channels.

*Comparison to the state-of-the-art*

Few studies have addressed top-down milling (TDM) of glass using femtosecond laser bursts [28,37,38]. In general, they report a substantial improvement in milling efficiency compared to the standard single-pulse modes, which can be seen in Figure 8(a). For the femtosecond pulse TDM, single-pulse mode yielded efficiencies of 6.8 and 10.1 $\mu m^3/\mu J$ for fused silica and soda-lime, respectively [37,38]. The use of GHz bursts with 10–15 pulses increased the TDM efficiency of fused silica by approximately a factor of seven, reaching ~ 50 μm³/μJ [28,38]. Further enhancement to 153 μm³/μJ was achieved by employing the BiBurst mode (P30N5), in which packets consisting of 30 pulses repeating at an intra-burst repetition rate of 2.1 GHz (GHz bursts) are repeated 5 times at an intra-burst repetition rate of 63.3 MHz (MHz bursts) [28].

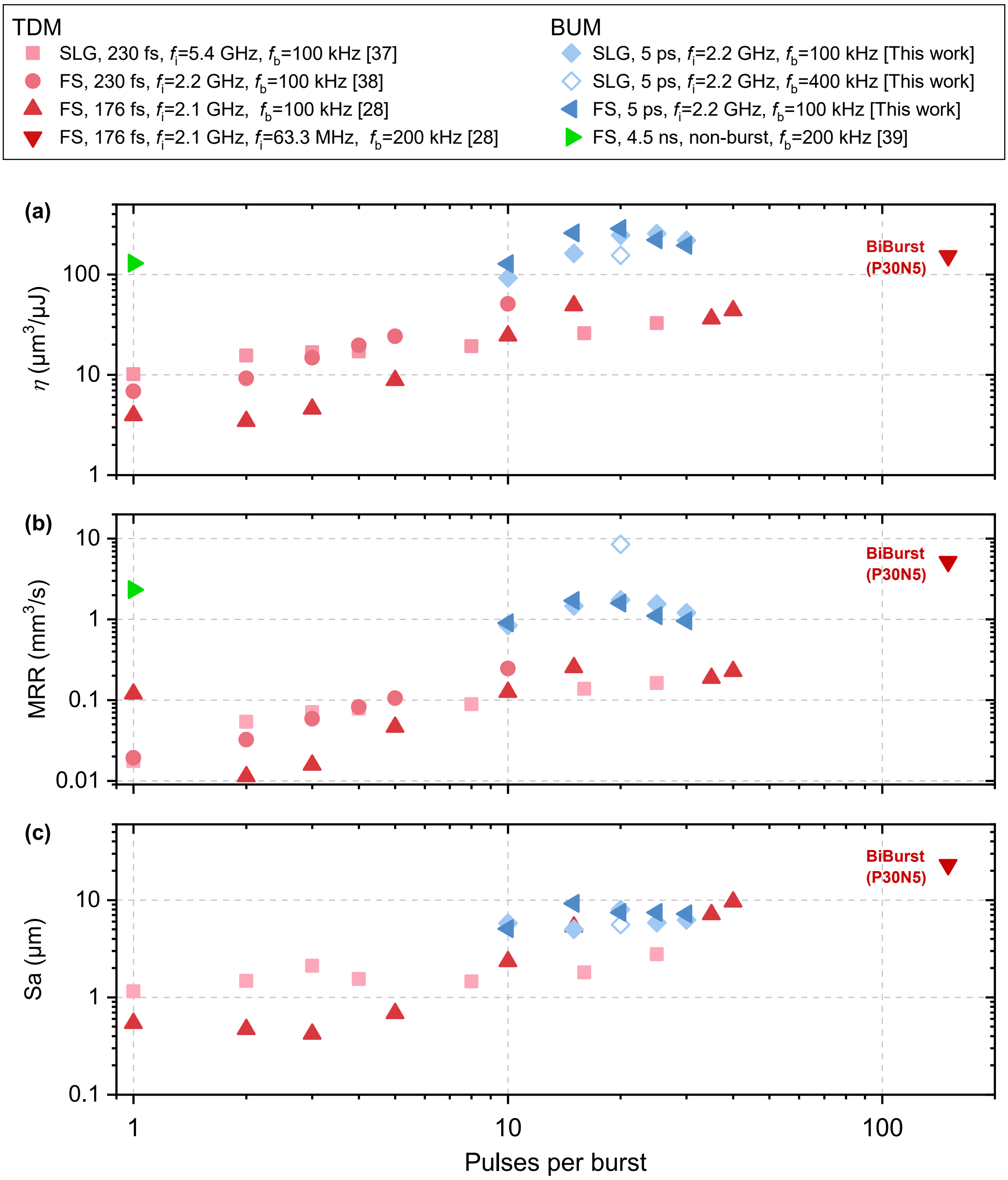

Figure 8. Comparison to the state of the art. **(a)** Milling efficiency, **(b)** material removal rate, and **(c)** processed surface average roughness Sa dependence on the number of pulses per burst for soda-lime glass (SLG) and fused silica (FS). Solid symbols denote conditions corresponding to the optimized milling efficiency, whereas open symbol indicate conditions corresponding to the optimized milling rate. All experiments were performed using a wavelength of 1 μm, except for Ref. [39], in which a wavelength of 532 nm was used.

BUM efficiency of 130 μm³/μJ has been reported for single-pulse operation using nanosecond pulses at a wavelength of 532 nm [39]. In the present work, a twofold increase in milling efficiency was achieved by employing a wavelength of 1030 nm and GHz bursts consisting of 20 pulses with picosecond duration. Material removal rates of up to 8.5 mm³/s were achieved with a 20-pulse GHz burst regime by increasing the burst repetition rate. Such a throughput was 1.6 times higher than that achieved with TDM using the BiBurst regime P30N5 [Figure 8(b)]. In contrast, material removal rates for all other TDM regimes remained below 0.25 mm³/s, whereas BUM employing nanosecond pulses or GHz burst regimes consistently reached several mm³/s. Overall, BUM provides superior milling efficiency and throughput compared to TDM, while the use of a GHz burst regime enhances the milling performance of both processing methods.

It should be noted that the taper angle of the cavities machined with TDM varied from 16 to 67°, with the most efficient processing regime yielding a taper angle of 31° [28]. In comparison, the BUM technique reaches higher milling efficiencies while maintaining vertical sidewalls (0° taper angle) (Figure 5).

Studies on TDM glass milling with GHz bursts indicate that the surface roughness generally increases with the number of pulses per GHz burst [Figure 8(c)]. The Sa is shown to increase from 0.4 to 9.6 μm when the number of pulses in the GHz burst increases from 3 to 40 [28]. This work does not indicate such a pronounced impact of the number of pulses per burst on surface roughness. The Sa ranged between 5.0 and 9.2 μm for 10–30 pulses in a GHz burst and was 5.6 μm under the highest-throughput regime ($f_b$ = 400 kHz). The results demonstrate that, under high-throughput operating conditions, BUM outperformed TDM by achieving material removal rates of up to 1.6 times higher while maintaining surface roughness approximately 4 times lower.

## 4. Conclusion

In this study, we demonstrate highly efficient and rapid bottom-up milling (BUM) of both fused silica and soda-lime glass. The latter was easier to process compared to fused silica, as no pre-roughening step was required. Our results show that the highest BUM efficiencies are achieved when using a large number of pulses per GHz burst (≥ 20). To the best of our knowledge, the milling efficiencies and rates reported here represent the highest values demonstrated for glass to date. For soda-lime glass, a milling efficiency of (256 ± 2) μm³/μJ [or (15.4 ± 0.1) mm³/min/W] was achieved, with a surface roughness Sa of (5.6 ± 0.2) μm and surface chippings of (56 ± 9) μm. For fused silica, a milling efficiency of (288 ± 2) μm³/μJ [or (17.3 ± 0.1) mm³/min/W] was obtained, with Sa of (7.5 ± 0.4) μm and surface chippings of (54 ± 6) μm. Furthermore, increasing the burst repetition rate to 400 kHz and maintaining a parameter set optimized for high milling efficiency enabled a soda-lime glass milling rate of (8.49 ± 0.03) mm³/s while maintaining a moderate Sa of (5.6 ± 0.5) μm. In addition to high throughput, the bottom-up strategy allows taper-less processing, which is not possible with the standard top-down approach.

The GHz burst regime enables high-throughput glass processing, while fewer pulses per burst, including the single-pulse regime, are known to provide high-quality machining. Therefore, tuning the burst parameters offers a promising route to combine high throughput with high machining quality and to selectively control the processed features. Furthermore, the same femtosecond laser source could be applied for other types of glass processing, such as volumetric modifications or surface structuring.

**Acknowledgements**

The authors gratefully thank Light Conversion for the laser source and Dr. Julijanas Željudevičius for his help with the temporal characteristic measurements.

## Supplementary material

The temporal characteristics of the GHz burst regimes are presented in Figure S1. A distinct pre-burst emerges for GHz burst regimes consisting of 20 and more pulses. The last pulse in the pre-burst occurred around 16 ns before the last pulse in the burst. The number of pre-burst pulses increased with the total number of pulses per burst, reaching a maximum for the $N = 30$ regime.

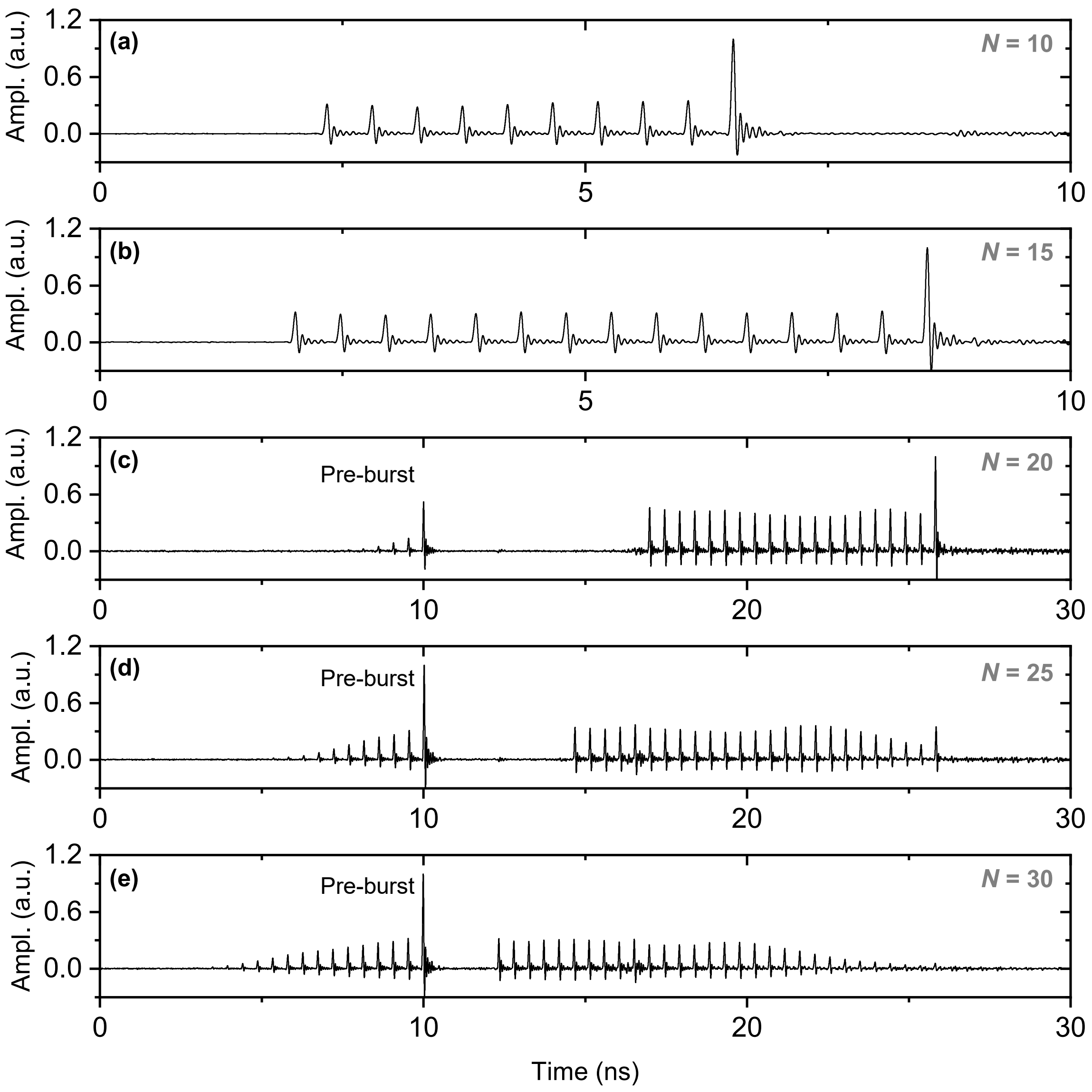

Figure S1. Temporal characteristics of GHz bursts. Oscillograms for regimes consisting of **(a)** 10, **(b)** 15, **(c)** 20, **(d)** 25, and **(e)** 30 pulses per GHz burst.

Surface roughness (Sa) and average chipping length dependence on milling efficiency is presented in Figure S2. For both glasses, surface roughness generally increased with milling efficiency. Conversely, the lowest roughness values (Sa < 4 µm) are obtained at the lowest milling efficiencies and the smallest number of pulses per burst investigated. The dependence of chipping length on milling efficiency is less pronounced. Nevertheless, the lowest number of pulses per burst also resulted in the smallest chippings, suggesting that this parameter plays an important role in limiting material damage.

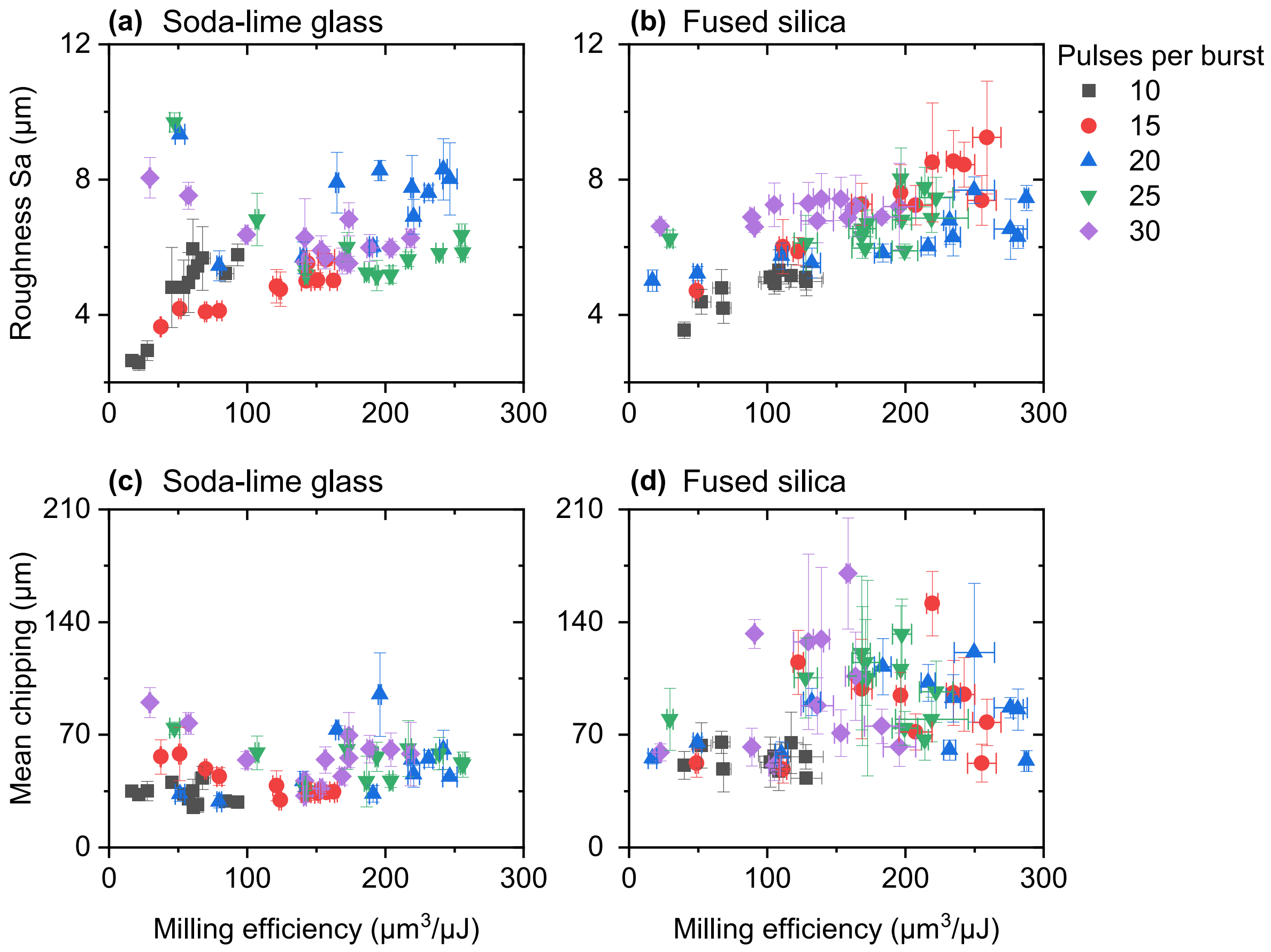

Figure S2. Quality versus milling efficiency. **(a, b)** Roughness Sa and **(c, d)** average chipping length of **(a, c)** soda-lime glass and **(b, d)** fused silica versus milling efficiency for 10, 15, 20, 25, and 30 pulses per GHz burst. The error bars along the y-axis represent the standard deviation calculated from 3–4 cavities measured under stable (linear) milling conditions. The error bars along the x-axis were calculated from the standard errors of the fitted slopes obtained from linear regression of cavity depth as a function of the number of processed layers.